\documentclass[conference]{IEEEtran}
\IEEEoverridecommandlockouts
\usepackage{cite}
\usepackage{amsmath,amssymb,amsfonts}
\usepackage{algorithmic}
\usepackage{graphicx}
\usepackage{epstopdf}
\usepackage{textcomp}
\usepackage{xcolor}
\usepackage{bm}
\usepackage{mathrsfs}
\usepackage{hyperref}
\usepackage[ruled,linesnumbered]{algorithm2e}
\usepackage{cuted}
\usepackage{subfigure}

\def\BibTeX{{\rm B\kern-.05em{\sc i\kern-.025em b}\kern-.08em
    T\kern-.1667em\lower.7ex\hbox{E}\kern-.125emX}}

\makeatletter
\newcommand{\linebreakand}{%
  \end{@IEEEauthorhalign}
  \hfill\mbox{}\par
  \mbox{}\hfill\begin{@IEEEauthorhalign}
}
\makeatother

\begin{document}

\title{Unified Near-field and Far-field Localization with Holographic MIMO}

\author{\IEEEauthorblockN{1\textsuperscript{st} Mengyuan Cao}
\IEEEauthorblockA{\textit{School of EECS} \\
\textit{Peking University}\\
Beijing, China \\
caomengyuan@pku.edu.cn}
\and
\IEEEauthorblockN{2\textsuperscript{nd} Haobo Zhang}
\IEEEauthorblockA{\textit{School of Electronics} \\
\textit{Peking University}\\
Beijing, China \\
haobo.zhang@pku.edu.cn}
\and
\IEEEauthorblockN{3\textsuperscript{rd} Hongliang Zhang}
\IEEEauthorblockA{\textit{School of Electronics} \\
\textit{Peking University}\\
Beijing, China \\
hongliang.zhang92@gmail.com}
\and
\IEEEauthorblockN{4\textsuperscript{th} Boya Di}
\IEEEauthorblockA{\textit{School of Electronics} \\
\textit{Peking University}\\
Beijing, China \\
diboya@pku.edu.cn}
}

\author{\IEEEauthorblockN{Mengyuan~Cao\IEEEauthorrefmark{1},
Haobo~Zhang\IEEEauthorrefmark{2},
Boya~Di\IEEEauthorrefmark{2},
and~Hongliang~Zhang\IEEEauthorrefmark{2}}
\IEEEauthorblockA{\IEEEauthorrefmark{1}School of Electronics Engineering and Computer Science, Peking University, Beijing, China.}
\IEEEauthorblockA{\IEEEauthorrefmark{2}School of Electronics, Peking University, Beijing, China.}
Email: \{caomengyuan, haobo.zhang, diboya\}@pku.edu.cn, hongliang.zhang92@gmail.com
}

\maketitle

\begin{abstract}
	Localization which uses holographic multiple input multiple output surface such as reconfigurable intelligent surface (RIS) has gained increasing attention due to its ability to accurately localize users in non-line-of-sight conditions. However, existing RIS-enabled localization methods assume the users at either the near-field (NF) or the far-field (FF) region, which results in high complexity or low localization accuracy, respectively, when they are applied in the whole area. In this paper, a unified NF and FF localization method is proposed for the RIS-enabled localization system to overcome the above issue. Specifically, the NF and FF regions are both divided into grids. The RIS reflects the signals from the user to the base station~(BS), and then the BS uses the received signals to determine the grid where the user is located. Compared with existing NF- or FF-only schemes, the design of the location estimation method and the RIS phase shift optimization algorithm is more challenging because they are based on a hybrid NF and FF model. To tackle these challenges, we formulate the optimization problems for location estimation and RIS phase shifts, and design two algorithms to effectively solve the formulated problems, respectively. The effectiveness of the proposed method is verified through simulations.
\end{abstract}


\section{Introduction}
A large number of emerging location-based applications, such as navigation, user tracking, and autonomous driving, have brought an urgent demand for high-precision localization. Such a demand has sparked research interest in developing advanced localization methods. Among various emergent techniques, localization using holographic multiple input multiple output (HMIMO) has received widespread attention due to its ability to achieve high localization accuracy. The HMIMO is facilitated by extremely large and near spatial continuous surfaces, which are integrated with numerous antennas or reconfigurable metamaterials. Such surfaces are capable of manipulating electromagnetic (EM) fields with high precision, thus providing high localization gain with a small size\cite{b32}\cite{b33}.

The passive HMIMO, also known as reconfigurable intelligent surface (RIS) is very popular due to its low power consumption and hardware costs. Specifically, the RIS is a planar surface consisting of many low-cost reflecting elements. It can shape the EM waves by varying the phase shifts of the elements, thus creating a customized signal path from the base station~(BS) to the user via reflection\cite{b30}. As the capability to customize the radio environment increases with the aperture of the RIS, a large RIS can be used to overcome the difficulty of lacking the line-of-sight (LOS) path and significantly enhance localization accuracy. 

Due to the large size of the RIS, the near-field (NF) region of the RIS is expanded\cite{b5}, where the spherical wave model is applied to characterize the signal propagation. In contrast, traditional localization methods primarily consider the localization in the far-field~(FF) region and utilize the plane wave model, which suffers low accuracy in the NF region. Recently, there has been a growing interest in addressing the NF localization probelm, which is based on the spherical wave model. In \cite{b24}, the authors examined a SISO scenario with an RIS, and resolved the localization problem using maximum likelihood. However, NF localization methods are not ideal for FF users, as the spherical wave model is more complex than the plane wave model, resulting in high complexity. Hence, a unified localization scheme is required to adapt to both the NF and FF regions.

In this paper, we propose a novel RIS-enabled unified NF and FF localization method to overcome the limitations of existing methods. In the considered scenario, a single-antenna user emits signals reflected by the RIS, and the single-antenna base station (BS) compares the received signals with the predicted signals at the sampled locations to locate the user. By optimizing the phase shifts of the RIS, the signals at different locations can be customized to improve the localization accuracy.

Several challenges need to be addressed for the proposed localization scheme. First, the design of the localization algorithm is complex because the signal models are different in the NF and FF regions. This complexity is further compounded by the uncertainty of whether the user is located in the NF or FF region. Second, optimizing the phase shifts of the RIS presents a challenge. This is primarily due to the enormous number of RIS elements and the non-convex nature of the optimization problem which stems from the constant modulus constraint of the RIS phase shifts.

Therefore, the main contributions of this paper can be summarized in the following:

1) We consider a RIS-enabled unified NF and FF localization scenario and introduce a localization protocol to coordinate the operations of the BS, the RIS, and the user during the localization process.

2) We formulate the location estimation problem that minimizes the localization loss and design a localization algorithm based on grid search to solve the formulated problem. We also formulate the RIS phase shift optimization problem to minimize the weighted Cramér-Rao bound (CRB) of the range and angles estimation errors. Then we propose a complex circle manifold-based method to obtain the suitable RIS phase shifts.

3) We compare the simulation results obtained by the proposed scheme and other RIS-enabled localization methods. It shows that our proposed method outperforms other methods in both the NF and FF regions given the same size of the RIS.

\begin{figure}
	\centering
	\includegraphics[scale=0.23]{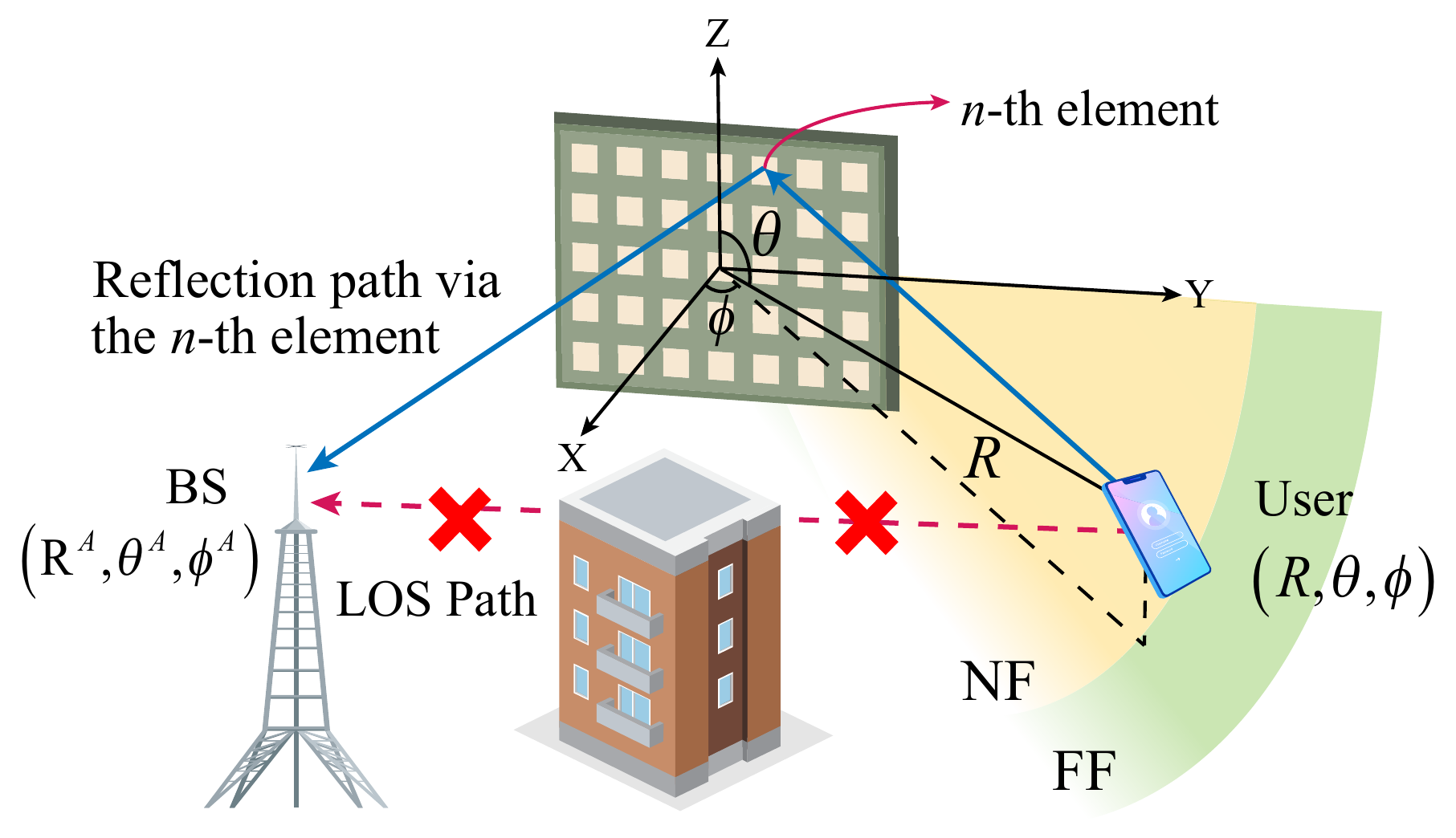}
	\caption{RIS-enabled NF and FF localization model.}
	\vspace{-0.3cm}
\end{figure}

The rest of the paper is organized as follows. In Sec. \ref{section:sm}, we provide the localization scenario, the received signal model, and the localization protocol. The localization algorithm is proposed in Sec. \ref{section:location}. In Sec. \ref{section:prob}, a RIS phase shift optimization problem is formulated, and an optimization algorithm is designed to solve the problem. In Sec. \ref{section:simu}, we present the simulation results. Finally, conclusions are drawn in Sec. \ref{sec:con}.

\section{System Model}
\label{section:sm}
In this section, we first introduce the scenario of the user localization, and then describe the signal model and the localization protocol for the proposed scenario.
\subsection{Localization Scenario}

As shown in Fig. 1., we consider a localization scenario consisting of a user, an RIS, and a BS. The user/BS is equipped with a single antenna. The RIS containing $N=L\times M$ elements is placed on the Y-Z plane, where the center of the RIS is at the origin of the coordinate system. We assume the line-of-sight (LOS) path between the user and the BS is blocked, and there only exists the reflection path via the RIS between the user and the BS.

During the localization process, the user sends a single-tone signal over frequency $f_c$ (wavelength $\lambda$) to the RIS, and the RIS reflects it to the BS. The BS then estimates the location of the user by using the received signals. We assume the locations of the RIS and the BS are known, and whether the user is located in the NF or the FF regions of the RIS is unknown.

\subsection{Signal Model}

In the uplink transmission, the signal $y$ received by the BS can be given by
\begin{equation}
y=\bm{\beta}^T\bm{h}(\bm{p})s+\epsilon,
\end{equation}
where $\bm{\beta}=[\beta_1,\beta_2,...,\beta_N]^T \in \mathbb{C}^{N\times 1}$ is the phase shift vector of the RIS, with $\beta_n$ representing the phase shift of the $n$-th RIS elements, and we have $|\beta_n|=1$. $s$ represents the narrow band signal transmitted by the user, and $\epsilon \sim \mathcal{CN}(0,\sigma^2)$ represents the received Gaussian noise at the BS with $\sigma^2$ being the noise power. $\bm{p}=[R,\theta, \phi]^T$ denotes the location of the user, with $R$, $\theta$, $\phi$ being the range, polar angle, and azimuth angle in the spherical coordinate system, respectively. The cascaded channel between the user and the BS $\bm{h}(\bm{p})$ is equal to $\bm{h}^A \odot \bm{h}^t(\bm{p})$, where $\bm{h}^A$, $\odot$, and $\bm{h}^t(\bm{p})$ denote the RIS-BS channel, Hadamard product and the user-RIS channel, respectively. 
Since the user could be in either NF or FF regions, the user-RIS channel models differ accordingly.

\subsubsection{Channel Model for the NF Region}
\label{subsubsection:nf}
When the user is located in the NF region, the transmitted signal should be described using spherical wave model and cannot be approximated as a plane wave for the RIS. Thus, the channel between the user and $n$-th RIS element is given by\cite{b6}
\setlength{\abovedisplayskip}{5pt}
\begin{align}
	h_{n}^t(\bm{p}) = \dfrac{\sqrt{G_t}\lambda}{4\pi d_{n}^t}\exp(-j\dfrac{2\pi}{\lambda}d_{n}^t),\  \bm{p} \in \bm{D}_{NF},
\end{align}
where $G_t$ is the transmit antenna gain, $d_{n}^t$ is the distance between the user and the $n$-th RIS element, $\bm{D}_{NF}$ represents the NF region. Based on \cite{b5}, the NF region is defined as the region where the maximum phase error between the phase shift calculated under plane wave approximation and the real phase shift is more than $\pi/8$.  Mathematically, the NF region is given by $ \bm{D}_{NF}=\{\bm{p} \vert \Delta \varphi(\bm{p}) > {\pi}/{8} \}$,
where $\Delta \varphi(\bm{p})$ is the maximum phase error across all the RIS elements, given by \par
\begin{small}
	\setlength{\abovedisplayskip}{-2pt}
	\begin{align}
		\Delta \varphi(\bm{p}) = \mathop{\max}\limits_{n}\frac{2\pi}{\lambda}\big(d_{n}^t - (R-y_n \sin\theta \sin\phi -z_n \cos\theta)\big),
	\end{align}
\end{small}%
where $(0, y_n, z_n)$ is the coordinate of the $n$-th RIS element.

\subsubsection{Channel Model for the FF Region}
\label{subsubsection:ff}
When the user is located in the FF region, we can adopt plane wave approximation to model the transmitted signals. Specifically, the channel gain is assumed to be the same for all the RIS elements, and the phase shifts adopt first-order approximation.
The channel between the $n$-th RIS element and the user in the FF region is~\cite{b6} \par
\begin{footnotesize}
	\setlength{\abovedisplayskip}{-2pt}
	\begin{align}
		h_{n}^t(\bm{p}) = &\dfrac{\sqrt{G_t}\lambda}{4\pi R}\exp (-j\dfrac{2\pi}{\lambda}(R-y_n\sin\theta \sin\phi -z_n \cos\theta) ), \bm{p} \in \bm{D}_{FF},
	\end{align}
\end{footnotesize}%
where $\bm{D}_{FF}$ represents the FF region. Based on \cite{b5}, the FF region is defined as the region where the maximum phase error is less than $\pi/8$. Mathematically, the FF region is given by $\bm{D}_{FF}=\{\bm{p}\vert \Delta \varphi(\bm{p}) < \pi/8 \}$.

\subsection{Localization Protocol}
\label{subsection:LP}
\begin{figure}
	\centering
	\setlength{\abovecaptionskip}{-0.1cm}
	\includegraphics[scale=0.31]{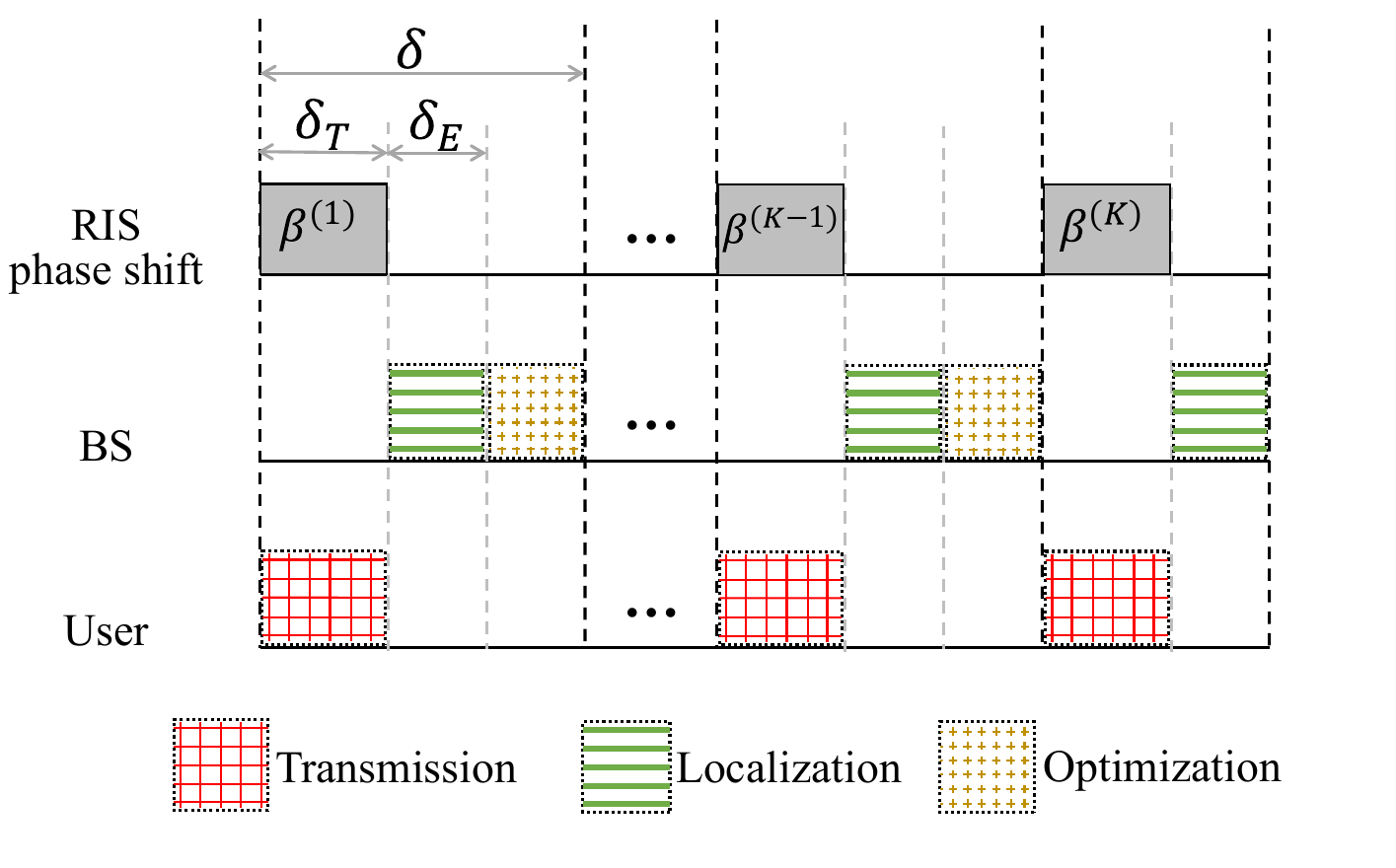}
	\caption{RIS-enabled mixed NF and FF source localization protocol.}
	\vspace{-0.35cm}
	\label{fig:2}
\end{figure}

In this subsection, we propose an RIS-enabled unified NF and FF source localization protocol, which improves localization accuracy by adaptively optimizing the RIS phase shifts based on the estimated location from the previous cycle. The localization process is divided into $K$ cycles with cycle duration $\delta$. Each cycle contains three steps: transmission, localization, and optimization. The process of the localization protocol is illustrated in Fig. \ref{fig:2}.

\subsubsection{Transimission}
In this step, the user sends a signal to the RIS for $\delta_T$ seconds, which is reflected to the BS. Let $g^{(k)}$ denote the signal received in the transmission step of the $k$-th cycle. The RIS phase shifts are set randomly in the first cycle, while in the following $K-1$ cycles, the phase shifts are selected based on the optimization results in the previous cycle, which will be described in the optimization step.

\subsubsection{Localization}
\label{subsubsection:localization}
In the next $\delta_E$ seconds, the BS estimates the user's location using the received signals. Specifically, in the $k$-th cycle, based on the received signals $\bm{g}=[g^{(1)},...,g^{(k)}]^T$, we determine whether the user is located in the NF or the FF region and estimate their location $\hat{\bm{p}}^{(k)}$. The details of the localization algorithm are introduced in Sec.~\ref{section:location}.

\subsubsection{Optimization}

In the rest time of the $k$-th cycle, the optimal RIS phase shifts $\bm{\beta}^{(k+1)}$ are selected according to the estimated location $\hat{\bm{p}}^{(k)}$, which are used to adjust the phase shifts of the RIS. Note that this step is not executed in the last cycle. The details of optimization are introduced in Sec.~\ref{section:prob}.

\section{RIS-enabled Unified Near-field and Far-field Localization}
\label{section:location}

In this section, we first formulate the localization problem and then propose a localization algorithm to solve the formulated problem.

\subsection{Localization Problem Formulation}
\label{sec:3A}
We formulate the localization problem to minimize the localization loss, which is defined as the $l_2$-norm of the residual between the received signal $\bm{g}$ and the signal reconstructed by using the estimated location. Thus, we can formulate the localization problem as
\begin{subequations}
	\begin{align}
		\label{prob:1}
			\text{P1}: &\mathop{\min}_{\bm{p}} \Vert \bm{g}-\bm{\beta}^T\bm{h}^A \odot \bm{h}^t(\bm{p})s \Vert_2^2 , \\
			&s.t. \ \ \bm{p} \in D_{NF} \cup D_{FF}.
	\end{align}
\end{subequations}

The objective function (\ref{prob:1}) in the above problem is non-convex, and thus the solution of (P1) easily falls into local minima by using conventional algorithms such as gradient descent. To solve it effectively, we propose a location estimation algorithm by modifying the grid search method in order to perform global search. Specifically, we sample the search domain and create a grid map of the sampled candidate locations. Note that for NF region, $R, \theta, \phi$ are all sampled because the range and angles are coupled in the NF region. While for FF region, the range $R$ is not sampled because the range only affects the overall channel amplitude, which can be addressed by linear regression. Besides, unlike the NF case, the range cannot be estimated for the FF user because the angles of arrival at all RIS elements are approximately the same. In this way, we can decrease the number of candidate locations and then decrease the algorithm complexity. The localization loss is calculated for each candidate location, and the one with the minimum loss is selected as the estimated location. 

\begin{figure}
	\centering
	\setlength{\abovecaptionskip}{0.cm}
	\setlength{\belowcaptionskip}{-50mm}
	\includegraphics[scale=0.23]{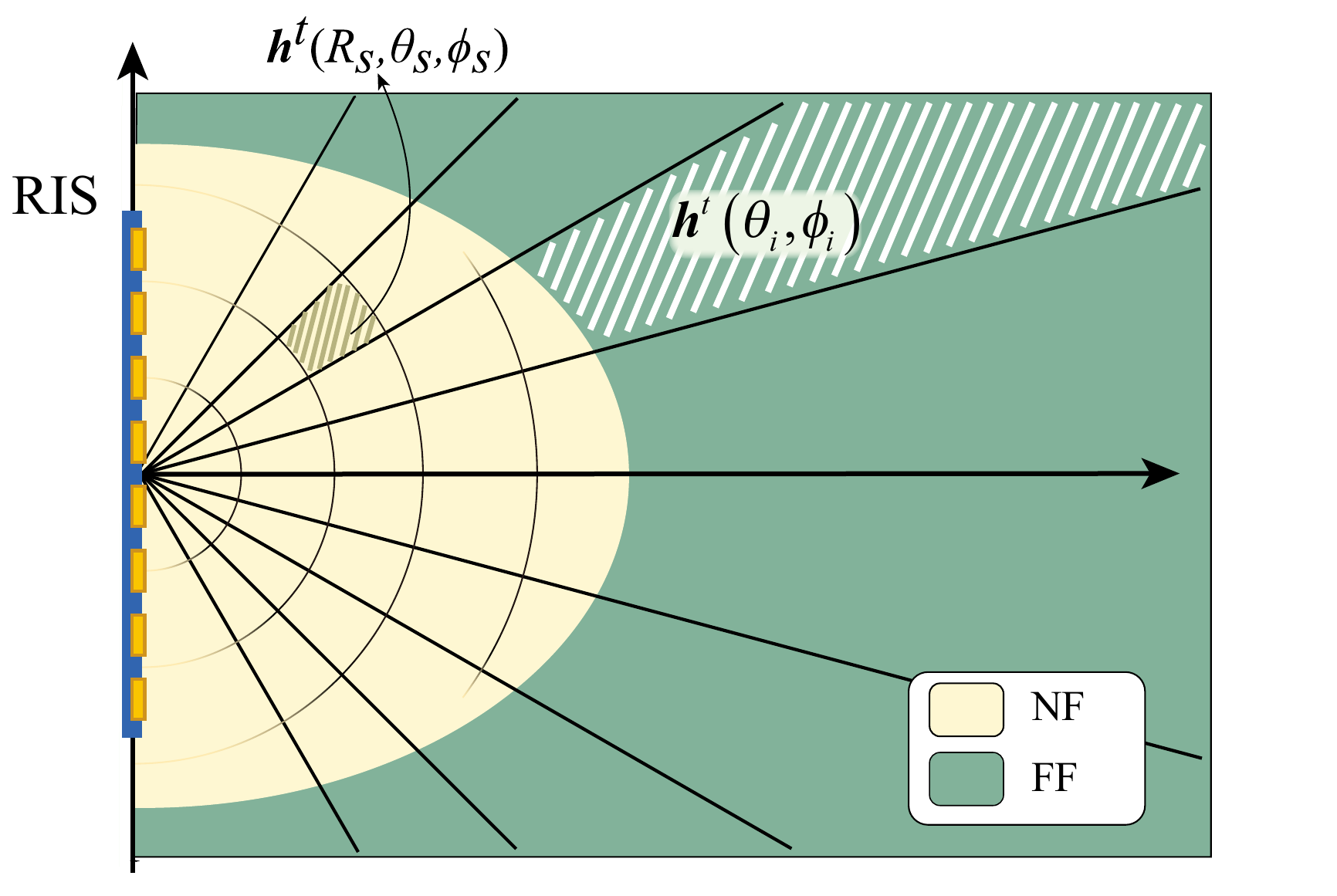}
	\caption{Space sampling method.}
	\vspace{-0.2cm}
	\label{fig:3}
	\vspace{-0.3cm}
\end{figure}

We define $\bm{F}=[\bm{F}_{near},\bm{F}_{far}]$ as the atom channels, which are the user-RIS channels between the RIS and every possible candidate locations of the user. $\bm{F}_{near}$ and $\bm{F}_{far}$ are the atom channels in the NF and FF regions, which will be explained in detail in the following. We also define $\bm{u}(\bm{p})$ as the coefficient matrix of atom channels. Since we only consider single user localization, only $1$ element in $\bm{u}$ should be non-zero. The cascaded channel is given by $\bm{h}(\bm{p})= \bm{h}^A \odot \bm{h}^t(\bm{p}) = \bm{h}^A \odot \bm{F}\bm{u}(\bm{p}) =diag\{\bm{h}^A\}\bm{F}\bm{u}(\bm{p})$. Hence, (P1) can be reformulated as
{
	\setlength\abovedisplayskip{6pt}
	\setlength\belowdisplayskip{6pt}
	\begin{subequations}
		\label{eq:6}
		\begin{align}
		\text{P1'}: \mathop{\min}_{\bm{u}} &\Vert \bm{g}-\bm{\beta}^Tdiag\{\bm{h}^A\}\bm{F}\bm{u} s \Vert_2^2 , \\
		&s.t. \Vert \bm{u}\Vert_0 =1.
		\end{align}	
	\end{subequations}
}

The atom channels in the NF region $\bm{F}_{near}$ is given by\cite{b13}
{
	\setlength\abovedisplayskip{7pt}
	\setlength\belowdisplayskip{7pt}
	\begin{equation}
		\bm{F}_{near} = [\bm{h}^t(R_1,\theta_1,\phi_1),...,\bm{h}^t(R_S,\theta_S,\phi_S)] \in \mathbb{C}^{N\times S} ,
	\end{equation}
}%
where $\bm{h}^t(R_s,\theta_s,\phi_s)\in \mathbb{C}^{N\times 1}$ is the user-RIS channel supposing that the user is located at $[R_s,\theta_s,\phi_s]^T  \in \bm{D}_{NF} $. Here, $S$ is the number of sampled locations in the NF, which is determined by the sampling spacings $\Delta R,\Delta \theta, \Delta \phi$. The range is uniformly sampled in the NF region with a constant spacing $\Delta R$. The polar and azimuth angles are uniformly sampled with the spacing $\Delta\theta=\pi/N_{\theta}$, $\Delta\phi=\pi/N_{\phi}$, where $N_{\theta}$ and $N_{\phi}$ are predetermined parameters. 

The atom channels of the FF region $\bm{F}_{far}$ is given by \cite{b12}
{
	\setlength\abovedisplayskip{7pt}
	\setlength\belowdisplayskip{7pt}
	\begin{align}
		\bm{F}_{far} = [\bm{h}^t(R_0,\theta_1,\phi_1),...,\bm{h}^t(R_0,\theta_{N_{\theta}N_{\phi}},\phi_{N_{\theta}N_{\phi}})]
	\end{align}
}%
where $\bm{h}^t(R_0,\theta_i,\phi_i)$ is the user-RIS channel assuming that user is at the $[R_0,\theta_i,\phi_i]$ in the FF. $R_0$ is a reference FF range that satisfies the FF condition for all angles. The sampling methods of angles are the same as in the NF case. 

For better illustration, the sampled locations at $\theta = \pi/2$ plane for both the NF and FF regions are shown in Fig. \ref{fig:3}.

\subsection{Localization Algorithm}
We propose an algorithm based on the exhaustive search of the possible candidate locations to solve (P1'). The basic idea is to calculate the loss of all the candidate locations and select the location with the smallest localization loss as the estimated location. Since the range is not sampled in the FF, the amplitudes of the atom channels are uncertain. However, in the NF, the amplitudes of the atom channels are known, indicating the loss calculation methods for the NF and the FF regions are different, which will be introduced separately.

\subsubsection{NF case}
In the NF region, we can calculate the localization loss directly by substituting each coefficient $u_i(i \leq S)$ with $1$. The localization loss for the $i$-th candidate location in the NF region is given by 
\begin{align}
	\label{equation:labegin}
	L^{near}_i = \Vert \bm{g}-\bm{\beta}^Tdiag\{\bm{h}^A\}\bm{F}_i s \Vert, (i \leq S),
\end{align}
where $\bm{F}_i$ is the $i$-th column of the matrix $\bm{F}$, i.e. the $i$-th atom channel in the NF region.

\subsubsection{FF case}
Since the channel amplitude is unknown in the FF region, we need to first obtain the estimated amplitude by linear regression, and then calculate the localization loss. Mathematically, the estimated amplitude for the $i$-th candidate location is given by
\begin{align}
	a_i &= (\bm{A}^H_i\bm{A}_i)^{-1}\bm{A}^H_i\bm{g} ,
\end{align}
where $\bm{A} = \bm{\beta}^Tdiag\{\bm{h}^A\}\bm{F}s$,
and $\bm{A}_i$ is the $i$-th column of the matrix $\bm{A}$.
Then, the localization loss of the $i$-th candidate location can be expressed as
\begin{align}
	L^{far}_i = \Vert \bm{g}  - a_i A_i \Vert, (i > S).
\end{align}

\subsubsection{Estimated location}
We choose the location corresponding to the minimal localization loss as the estimated location, which is denoted by $i^*$. i.e. 
\begin{align}
	\label{equation:laend}
	i^* = \mathop{\arg\min}_i \{ L^{near}_i, L^{far}_i \}.
\end{align}
If $i^* \leq S$, the user is in the NF region, and otherwise, the user is estimated in the FF region, and the estimated location corresponds to $u_{i^*}$, i.e.,
\begin{align}
\nonumber
	&\hat{\bm{p}}^{(k)}=\left\{
	\begin{aligned}
	&[R_i,\theta_{i^*},\phi_{i^*}]^T,&         &{i^*} \leq S \ (\text{NF \ region}),\\
	&[\theta_{{i^*}-S},\phi_{{i^*}-S}]^T,&         &{i^*} > S \ (\text{FF \ region}).\\
\end{aligned}
\right.
\end{align}

\section{RIS Phase Shift Optimization}
\label{section:prob}

In this section, we first formulate the RIS phase shift optimization problem. The optimization problems for the NF and FF cases are formulated separately. This is because in the localization step of the protocol, whether the user is located in the NF or the FF region has been determined and the location parameters for the NF and FF cases are different. Then, an optimization algorithm based on complex circle manifold (CCM) is proposed to solve the formulated problems.

\subsection{Phase Shift Optimization Problem Formulation}

\subsubsection{Problem Formulation for the NF Case}
If the estimated location $\hat{\bm{p}}^{(k)}$ is in the NF region in the $k$-th cycle, we estimate range and angles in the $(k+1)$-th cycle, and the estimated parameters are denoted by $\bm{p}=[R,\theta, \phi]^T$.
Similar to the proof in \cite{b6}, for the NF case, the CRBs of the unknown parameters $\bm{p}=[R,\theta, \phi]^T$ can be given by
{
	\setlength\abovedisplayskip{8pt}
	\setlength\belowdisplayskip{8pt}
	\begin{align}
		\label{equation:nf_crb}
		D_R &=\left(\bm{J}_{NF}^{-1}\right)_{1,1},\\
		D_\theta &=\left(\bm{J}_{NF}^{-1}\right)_{2,2},\\
		D_\phi &=\left(\bm{J}_{NF}^{-1}\right)_{3,3},
	\end{align}
}%
where $\bm{J}_{NF}$ is the $3 \times 3$ Fisher information matrix of $\bm{p}$ in the NF region, and the $(i, j)$-th element in $\bm{J}_{NF}$ is given by
\begin{small}
\begin{equation}
[\bm{J}_{NF}]_{i,j} = \dfrac{2}{\sigma^2} \sum_{q=1}^{k+1}Re \left \{ \dfrac{\partial (\mu^{(q)})^H}{\partial p_i} \dfrac{\partial \mu^{(q)}}{\partial p_j} \right \},
\end{equation}
\end{small}%
where $\frac{\partial \mu^{(q)}}{\partial p_i}=(\bm{\beta}^{(q)})^T \frac{\partial \bm{h}^{NF}}{\partial p_i}s$. $\bm{h}^{NF}$ is the NF channel calculated by using $\hat{\bm{p}}^{(k)}$ and $p_i$ is the $i$-th element in $\bm{p}$. The $n$-th entry of $\frac{\partial \bm{h}^{NF}}{\partial p_i}$ can be expressed as
\begin{footnotesize}
	\setlength\abovedisplayskip{6pt}
	\setlength\belowdisplayskip{5pt}
	\begin{align}
	\left[ \dfrac{\partial \bm{h}^{NF}}{\partial p_i}\right]_{n} = h_{n}^A \dfrac{\sqrt{G_t}\lambda}{4\pi d_{n}^t} \dfrac{\partial}{\partial p_i}\left(\exp\left(-j\dfrac{2\pi}{\lambda}d_{n}^t \right)\right) = h_{n} \left(-j\dfrac{2\pi}{\lambda}\right)\dfrac{\partial d_{n}^t}{\partial p_i},
\end{align}\end{footnotesize}%
where
{
	\setlength\abovedisplayskip{8pt}
	\setlength\belowdisplayskip{8pt}
	\begin{small}
		\begin{align}
			\frac{\partial d_{n}^t}{\partial p_1} = \frac{\partial d_{n}^t}{\partial R} &= \frac{1}{d_{n}^t}(R- y_n\sin\theta \sin\phi -z_n\cos\theta ), 
			\\ 
			\frac{\partial d_{n}^t}{\partial p_2} = \frac{\partial d_{n}^t}{\partial \theta} &= \frac{1}{d_{n}^t}(-Ry_n\cos\theta \sin\phi  +Rz_n\sin\theta ), \\
			\frac{\partial d_{n}^t}{\partial p_3} = \frac{\partial d_{n}^t}{\partial \phi} &= \frac{1}{d_{n}^t}(-Ry_n\sin\theta \cos\phi ).
		\end{align}
	\end{small}%
}%

Thus, the optimization problem in the NF region can be written as
\begin{subequations}
	\begin{align}
		\text{P2}: & \mathop{\min}_{\bm{\beta}^{(k+1)}} g(\bm{p},\bm{\beta}^{(k+1)}) = tr \left(\bm{J}_{NF}^{-1}W^{NF} \right)\\
		& s.t. \quad |\beta_n ^{(k+1)}|=1, \forall n=1,2,...N, \label{eq:sub21b}
	\end{align}
\end{subequations}
where $\beta_n ^{(k+1)}$ is the phase shift of the $n$-th RIS element in the $(k+1)$-th cycle, and $W^{NF}=diag \{w_1,w_2,w_3 \}$ is the weight matrix of the CRBs in the NF region. 
Here $w_i$ is the predetermined weight of $p_i$.

\subsubsection{Problem Formulation for the FF Case}
If the estimated location in the $k$-th cycle $\hat{\bm{p}}^{(k)}$ is in the FF region, we only estimate the angles in the $(k+1)$-th cycle, and the estimated parameters are denoted by $\bm{p}=[\theta, \phi]^T$.
For the FF case, the CRBs of the unknown parameters $\bm{p}$ can be given by
\begin{align}
	D_\theta &=\left(\bm{J}_{FF}^{-1}\right)_{1,1},\\
	D_\phi &=\left(\bm{J}_{FF}^{-1}\right)_{2,2},
\end{align}
where $\bm{J}_{FF}$ is the $2\times 2$ Fisher information matrix of $\bm{p}$ in the FF region, and the $(i, j)$-th element is given by
\begin{small}
	\begin{equation}
		[\bm{J}_{FF}]_{i,j} = \dfrac{2}{\sigma^2} \sum_{q=1}^{k+1}Re \left \{ \dfrac{\partial (\mu^{(q)})^H}{\partial p_i} \dfrac{\partial \mu^{(q)}}{\partial p_j} \right \},
	\end{equation}
\end{small}%
where $\frac{\partial \mu^{(q)}}{\partial p_i}=(\bm{\beta}^{(q)})^T \frac{\partial \bm{h}^{FF}}{\partial p_i}s$.
and $\bm{h}^{FF}$ is the FF channel calculated using $\hat{\bm{p}}^{(k)}$.

In the FF region, we have
\begin{small}
	\begin{align}
		\left [\dfrac{\partial \bm{h}^{FF}}{\partial p_1}\right]_{n} = h_{n} \left(-j\dfrac{2\pi}{\lambda} \right)&(-y_n\cos\theta \sin\phi + z_n\sin\theta),\\
		\left [\dfrac{\partial \bm{h}^{FF}}{\partial p_2}\right]_{n} = h_{n} \left(-j\dfrac{2\pi}{\lambda} \right)&(-y_n \cos\theta \cos\phi).
	\end{align}
\end{small}%

Thus, the optimization problem in the FF region can be formulated as
\begin{subequations}
	\setlength{\belowdisplayskip}{3pt}
	\begin{align}
		\text{P3}: & \mathop{\min}_{\bm{\beta}^{(k+1)}} g(\bm{p},\bm{\beta}^{(k+1)}) = tr \left(\bm{J}_{FF}^{-1}W^{FF} \right)\\
		& s.t. \quad |\beta_n^{(k+1)}|=1, \forall n=1,2,...N, \label{eq:sub27b}
	\end{align}
\end{subequations}
where $W^{FF}=diag \{w_1,w_2 \}$ is the weight matrix of the CRBs in the FF region. 
Here $w_i$ is the preset weight of $p_i$.

\subsection{RIS Phase Shift Optimization Algorithm}

The objective function in (P2) contains the CRBs of range and angle, while that in (P3) only has the latter. Thus, (P3) can be viewed as a special case of (P2). In this subsection, we only introduce the algorithm for (P2) in the following and omit that for (P3) due to the page limit.

Specifically, we design an optimization algorithm base on CCM method to tackle (P2). Due to the constant modulus constraint (\ref{eq:sub21b}), the optimization problem (P2) is non-convex, which is numerically difficult to handle. Fortunately, the solution can be considered as lying on the CCM to satisfy the constant modulus, and the manifold can be represented as \par
\begin{small}
	\setlength{\abovedisplayskip}{-1pt}
	\begin{equation}
		\mathcal{M}^N= \left\{ \bm{\beta}^{(k+1)} \in \mathbb{C}^{N}:|\beta_1^{(k+1)}|=...=|\beta_N^{(k+1)}|=1 \right\}.
	\end{equation}
\end{small}%

The main idea of the CCM-based optimization method is to iteratively apply gradient descent in the manifold space. After several iterations, the algorithm terminates when the difference between two iterations of $g(\bm{\beta}^{(k+1)})$ is less than a constant $\zeta$.
The CCM-based optimization method is composed of four main steps in each iteration:

\subsubsection{Compute the gradient in Euclidean space}
We use the Euclidean gradient as the search direction for the minimization problem in Euclidean space. In the NF region, the Euclidean gradient of $f(\bm{\beta}^{(k+1)})$ is given by \par
\begin{small}
	\setlength{\abovedisplayskip}{0pt}
\begin{align}
\label{eq:nfeg}
\nabla g(\bm{\beta}^{(k+1)})
&=2 \left[ \dfrac{ w_1 \partial D_R}{\partial{\bm{\beta}^{(k+1)}}^{\ast}} + \dfrac{w_2 \partial D_\theta}{\partial{\bm{\beta}^{(k+1)}}^{\ast}} +  \dfrac{w_3 \partial D_\phi}{\partial{\bm{\beta}^{(k+1)}}^{\ast}} \right].
\end{align}
\end{small}%
Similar to the proof in \cite{b14}, the derivatives of CRBs in the NF region with respect to ${\bm{\beta}^{(k+1)}}^{\ast}$ can be derived.
\footnote{
	For FF case, the difference lies in the first step of the algorithm. Since the range of the FF user cannot be estimated, the first term in the Euclidean gradient is omitted.
}

\subsubsection{Compute the Riemannian gradient}
The Riemannian gradient is the projection of the gradient onto the tangent space of the complex circle manifold. The Riemannian gradient of the objective function $g(\bm{\beta}^{(k+1)})$ at the point $\bm{\beta}^{(k+1)}_j$ on the complex circle manifold $\mathcal{M}$ can be given as \cite{b14} \cite{b15}\par
\begin{small}
	\setlength{\abovedisplayskip}{-2pt}
	\setlength{\belowdisplayskip}{1pt}
	\begin{align}
	\label{eq:nfg}
	\nabla_{\mathcal{M}}g(\bm{\beta}^{(k+1)}_j&)= -\nabla g(\bm{\beta}^{(k+1)}_j) \nonumber \\
	&-Re \left\{ \left(\nabla g(\bm{\beta}^{(k+1)}_j)\right)^{*} \odot {\bm{\beta}^{(k+1)}_j} \right\} \odot \bm{\beta}^{(k+1)}_j.
	\end{align}
\end{small}

\subsubsection{Update over the Tangent Space}
We choose a step size to update the current point, which is mathematically given as
\begin{small}
\begin{align}
\label{equation:barj}
	\bar{\bm{\beta}}^{(k+1)}_j = \bm{\beta}^{(k+1)}_j  + \alpha_j \nabla_{\mathcal{M}}g(\bm{\beta}^{(k+1)}_j),
\end{align}
\end{small}%
where $\alpha_j$ is the step size in the $j$-th iteration.

\subsubsection{Retract onto the manifold}
After the update, the new point $\bar{\bm{\beta}}^{(k+1)}_j$ generally does not lie on the manifold $\mathcal{M}$. By using the retraction operator, the new point is mapped into the manifold. The retraction operator is given as
\begin{small}
\begin{align}
\label{equation:j1}
	\bm{\beta}^{(k+1)}_{j+1}= \bar{\bm{\beta}}^{(k+1)}_j \odot \dfrac{1}{ |\bar{\bm{\beta}}^{(k+1)}_j |}.
\end{align}
\end{small}

The proposed algorithm is summarized in \textbf{Algorithm~\ref{alg:algorithm1}}.

\setlength{\textfloatsep}{0.1cm}
\begin{algorithm}[t]
\label{alg:algorithm1}
	\caption{CCM-based RIS Phase Shift Optimization Algorithm}
	\KwIn{Estimated location $\hat{\bm{p}}^{(k)}$, RIS phase shift for previous $k$ cycles $\bm{\beta}^{(q)},q=1,2,...k$}
	\KwOut{Optimal RIS phase shift $\bm{\beta}^{(k+1)}$}  
	\BlankLine
	Initialize: $j=0$, $\beta_0 \in \mathcal{M}$; \\
	\While{\begin{small}$|g(\bm{\beta}^{(k+1)}_{j+1}) - g(\bm{\beta}^{(k+1)}_{j})| < \zeta $\end{small}} {
		\eIf{$\hat{\bm{p}}^{(k)} \in D_{NF}$}{
			Compute the Euclidean gradient \begin{small}$\nabla g(\bm{\beta}^{(k+1)}_j)$\end{small}%

			according to (\ref{eq:nfeg}); 
		}{
			Compute the Euclidean gradient \begin{small}$\nabla g(\bm{\beta}^{(k+1)}_j)$\end{small}%

			according to (\ref{eq:nfeg}) with the first term omitted; \\}
		Calculate the Riemannian gradient \begin{small}$\nabla_{\mathcal{M}}g(\bm{\beta}^{(k+1)}_j)$\end{small} according to (\ref{eq:nfg}); \\
		Compute the RIS phase shift update on the tangent space\begin{small}$\bar{\bm{\beta}}^{(k+1)}_j$\end{small} according to (\ref{equation:barj}); \\
		Update RIS phase shift \begin{small}$\bm{\beta}^{(k+1)}_{j+1}$\end{small} according to (\ref{equation:j1}); \\ 
		$j = j+1$;
	}
\end{algorithm}
\setlength{\floatsep}{0.1cm}

\begin{figure*}
	\begin{minipage}[t]{1\textwidth}
	\centering
	\setlength{\abovecaptionskip}{-0.15cm}
	\includegraphics[height=1.57in]{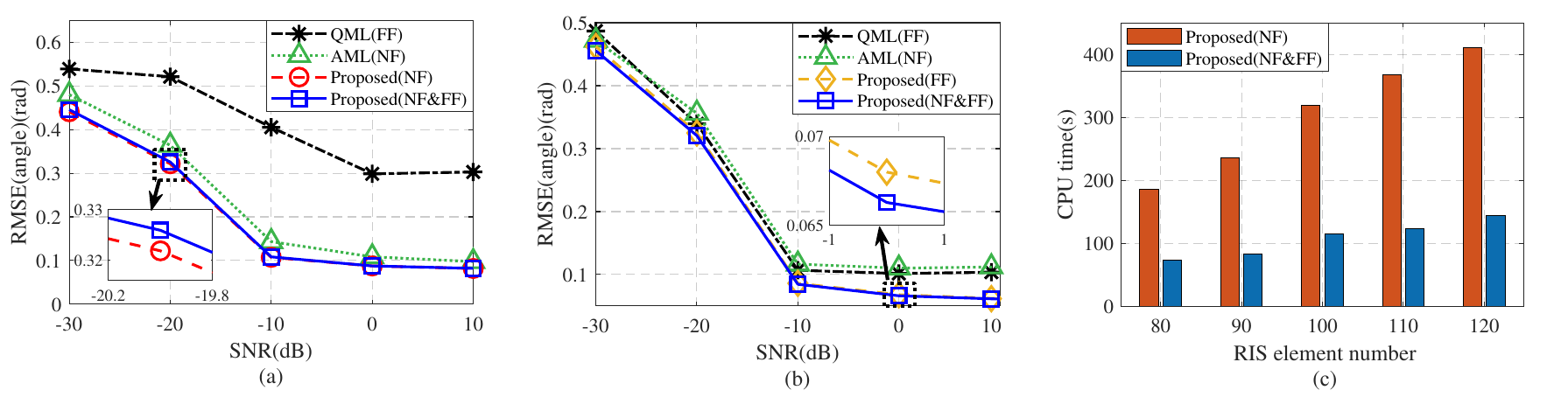}
	\caption{Angle estimation accuracy of the proposed algorithm compared 
	with QML and AML when the user is located (a) in the NF region and 
	(b) in the FF region. (c) Cputime of the proposed algorithm in unified NF and FF model and NF model.}
	\label{pic:angle}
	\end{minipage}%
	\vspace{-0.3cm}
\end{figure*}

\section{Simulation Results}
\label{section:simu}

This section presents the simulation results to demonstrate the performance of the proposed method. Simulations are conducted on a personal computer with a 1.80GHz Intel Core i7 processor.
We use an RIS with $10 \times 10$ elements. The element spacing is $\lambda/2$. The transmit frequency is 28GHz. 
The polar angle and the azimuth region are evenly divided into $10$ grids, and the range sampling spacing is $5\lambda$. We set the cycle number as $K=15$ and conduct $T = 500$ independent trials.

Fig. \ref{pic:angle}(a) and (b) show the RMSEs of angle estimation versus the signal-to-noise ratio (SNR) when the user is in the NF and the FF regions, respectively. The SNR is defined as $P_s / \sigma^2$\cite{b28}, where $P_s = ((\bm{h}^A)^T \bm{I} \bm{h}^t)^2$ is the signal power defined as the received signal power when the phase shifts of all the RIS elements are 1. For comparison, we also provided the results of an RIS-enabled NF localization method (AML in\cite{b20}) and an RIS-enabled FF localization method (QML in\cite{b19}). Besides, we combine the proposed algorithm with pure NF or FF models to show the gain brought by the proposed algorithm rather than the hybrid NF and FF models, which are labeled as Proposed (NF) and Proposed (FF), respectively. It can be seen that as SNR increases, the RSME first decreases and then remains fixed. The RMSE does not decrease when the SNR is sufficiently large because the estimated location is at one of the sampled grids, which may not be the actual location of the user. We can also observe from Fig. \ref{pic:angle}(a) and (b) that the proposed algorithm with the hybrid NF and FF model, labeled as Proposed (NF\&FF), outperforms the AML and QML algorithms in both the NF and FF cases, which shows the superiority of the proposed method. Besides, it is shown in Fig.~\ref{pic:angle}(a) that the RMSE obtained by the proposed algorithm applied in the NF model is slightly lower than the proposed algorithm with the NF\&FF model, while in Fig.~\ref{pic:angle}(b), the RMSE of the proposed algorithm applied in the FF model is a little bit higher. This is because the NF model has the highest accuracy when describing the wireless signals, while the accuracy of the FF model is the lowest.

Fig. \ref{pic:angle}(c) illustrates the CPU time versus RIS element number.
It shows that the proposed algorithm applied in the NF model has much higher CPU time than that in the hybrid NF and FF model, indicating that the proposed algorithm applied in the NF model has higher complexity. This is because given the same space of interest, the number of sampling locations in the NF model is much larger than that in the hybrid model, which increases the overall complexity. In addition, we can observe that as the RIS element number increases, the CPU time also increases. 
This is because the complexity of the optimization algorithm is positively correlated with the RIS 
element number.

\section{Conclusion}
\label{sec:con}
In this paper, we have developed a RIS-enabled unified NF and FF localization method. A unified NF and FF localization scenario has been considered and a localization protocol has been introduced to regulate the operations of the BS, the RIS, and the user. A localization algorithm based on grid search has been designed for location estimation. We have analyzed the CRBs of the estimated location parameters, and designed a CCM-based optimization algorithm, which has lowered the CRB to further improve the localization accuracy. Simulation results have shown that: 1) the proposed localization method provides better localization accuracy than the compared algorithms given the same size of the RIS; 2) the range estimation accuracy is not sensitive to the variation of the SNR, and a centimeter accuracy can still be achieved in low SNR conditions.



\end{document}